# Global Metallicities From Globulars Through To Elliptical Galaxies

K. RAKOS[1], J. SCHOMBERT[2], A. ODELL[3], M. MAITZEN[1]
*(1) Institute for Astronomy, University of Vienna*
*(2) Dept. of Physics, University of Oregon*
*(3) Dept. of Physics and Astronomy, Northern Arizona University*

**Abstract**

We present narrow band data on dwarf ellipticals in nearby (Virgo, Fornax) and intermediate redshift clusters (A2125, A2218) in order to study their mean age and metallicities. In all four clusters, nucleated dwarf ellipticals display colors that place them on the high metallicity end of the Milky Way globular cluster sequence. Normal dwarf ellipticals display colors, and resulting metallicities and ages, that align them with bright ellipticals. This suggests that dE,N's may be the ancestors to the red GC population found around many elliptical galaxies.

## 1.1 Introduction

A majority of stellar population studies have focused on the determination of age and metallicity through the use of various spectral signatures in the light from galaxies. An alternative approach to spectral line studies is to examine the shape of specific portions of a galaxy's SED using narrow band filters centered on specific regions sensitive to the RGB (metallicity) and the turnoff point (mean age) stars. For the last fifteen years, K. Rakos *et al.* have explored the behavior of galaxy colors as a function of redshift with a narrow band filter system (a modified Strömgren system). In numerous papers, we have shown following advantages of Strömgren photometry in extragalactic research. One, it is possible to use 'redshifted' filters corresponding to the redshift of the galaxy cluster in consideration instead of the use of uncertain k-corrections. The observations are always made in the rest system (Rakos, Fiala & Schombert 1988), thus, for each cluster a separate set of filters is used, Second, the use of a redshifted filter set allows for the determination of cluster membership without time consuming spectroscopy. A high precision method is described in Steindling, Brosch & Rakos (2001) and in Odell, Schombert & Rakos (2002). Lastly, being a blue/near-UV color system, the Strömgren colors provide a measure of recent star formation (integrated over that last Gyr) and can be used to photometrically classify galaxies into passive (elliptical), star forming (spirals) and starburst (irregulars) systems. (Rakos, Maindl & Schombert 1996).

Color indices in quiescent systems (e.g. ellipticals) are correlated to the metallicity and age of galaxies (Rakos *et al.* 2001 and Odell, Schombert & Rakos 2002). Objects composed of single stellar populations (SSP, i.e., globular clusters) or a composite of SSPs (i.e. ellipticals) present special circumstances for the study of the evolution of stellar populations. Any investigation into the stellar population in ellipticals must focus on the separation of the age and metallicity to the underlying stars as the two key parameters. The age-metallicity

*K. Rakos, J. Schombert, A. Odell, M. Maitzen*

Table 1.1. *Table 1. Dwarf Elliptical Colors*

| N | $uz-vz$ | $bz-yz$ | $vz-yz$ | $mz$ | [Fe/H] | Cluster | z |
|---|---|---|---|---|---|---|---|
| dE,N galaxies : | | | | | | | |
| 40 | 0.63 | 0.31 | 0.47 | -0.15 | -1.05 | Fornax | 0.01 |
| 42 | 0.62 | 0.32 | 0.47 | -0.17 | -1.05 | Coma | 0.02 |
| 40 | 0.50 | 0.29 | 0.49 | -0.09 | -0.98 | A2218 | 0.17 |
| 33 | 0.63 | 0.27 | 0.39 | -0.15 | -1.36 | A2125 | 0.24 |
| dE galaxies : | | | | | | | |
| 6 | 0.76 | 0.27 | 0.62 | 0.08 | -0.48 | Fornax | 0.01 |
| 14 | 0.62 | 0.30 | 0.69 | 0.09 | -0.22 | Coma | 0.02 |
| 18 | 0.76 | 0.31 | 0.69 | 0.07 | -0.22 | A2218 | 0.17 |
| 16 | 0.77 | 0.27 | 0.67 | 0.13 | -0.29 | A2125 | 0.24 |

degeneracy can be broken with the use of Strömgren color indices. For example, Bell and Gustafsson (1978) have shown that the metallicity is a function of the color indices [Fe/H] = $(m_1 + a_1(b-y) + a_2)/(a_3(b-y) + a_4)$. In addition, recently J. Schulz *et al.* (2002) has published Spectral & Photometric Evolution of Simple Stellar Populations at Various Metallicities in Gottingen. These models display a complete separation between [Fe/H] of −0.7 and −1.7 for different ages from 0.84 to 14 Gyr if using Strömgren colors.

## 1.2 Identifying Dwarf Ellipticals Through Their Continuum Colors:

Dwarfs observed in the Fornax galaxy cluster (Rakos *et al.* 2001) can be divided in two groups normal (dE) and nucleated (dE,N) according to the FCC classification. Strömgren photometry shows that dE,N are a direct continuation of globular clusters toward larger masses, in terms of their color indices. Whereas, dEs are more similar to their more massive cousins, the giant ellipticals. If galaxies are constructed by hierarchical mergers, then the dE's must be fragments of the original protogalaxies or fossils remains of the galaxy formation epoch.

In addition to the Fornax cluster, we have observed Coma, A2125 (a high blue fraction Butcher-Oemler cluster) and A2218 (one of the richest Abell clusters). The sample includes galaxies between M = −16 and −23 and uses the color indices of dEs and dE,Ns from Fornax photometry (classed by morphology) and selected dwarfs with similar luminosity and color indices in the more distant clusters. Table 1 displays the results. We see that A2218, a very dense cluster, has 3 times more dEs galaxies than in Fornax, while both have equal number of dE,Ns (observed under identical conditions). The number of dEs are evidently a product of the cluster density and therefore of a large number of galaxy interactions with time.

Rakos *et al.* (2001) has shown a very strong linear correlation between the color ($vz-yz$) and the metallicity in globular clusters, dwarf elliptical galaxies and in similar form an approximation also used for normal elliptical galaxies with the help of multi-metallicity models. Table 1 shows that luminosity-weighted [Fe/H] of dE,N galaxies depend on their ages, with A2125 at the highest redshift and the lowest metallicity. In contrast, A2218 is very dense cluster with high metallicity, and it may be that the cluster density is also correlated with metallicity.

*K. Rakos, J. Schombert, A. Odell, M. Maitzen*

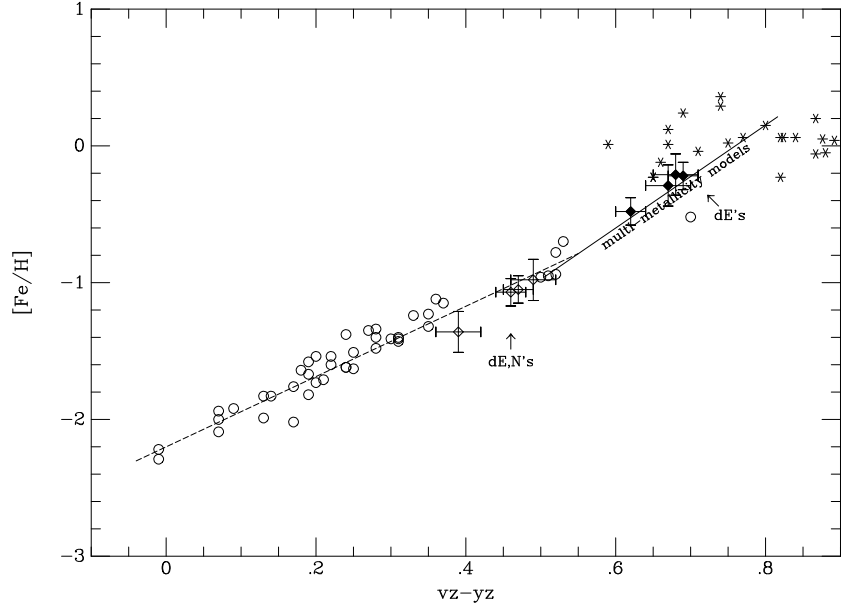

Fig. 1.1. Metallicity color versus [Fe/H] for MW globulars (open circles), ellipticals (stars), and dwarf ellipticals. dE's classified by morphology as nucleated distinguish themselves as being globular-like in their colors and metallicities compared to normal dE's.

Figure 1 shows the relationship between metallicity [Fe/H] and modified Strömgren color $vz-yz$ for globular clusters in our Galaxy (open circles, determined from fits to CMDs) as collected from the literature and in general have the accuracy of ±0.1 dex. Normal ellipticals are filled circles with metallicities from Trager *et al.* (2000). The $vz-yz$ color accuracy is ±0.02 magnitude or better and is based on our observations. Globular clusters show linear correlation for all colors $0 < vz-yz < 0.7$, the same is true for nucleated dwarf ellipticals for $0.4 < vz-yz < 0.7$. Normal elliptical galaxies show a larger real scatter but the dE galaxies join the same region of the diagram as normal E galaxies. The scatter of E galaxies is probably induced by cannibalism in the past, see Trager *et al.* (2000). For Globulars

$$[Fe/H] = -2.185 + 2.587(vz-yz)(R^2 = 0.95 \, for \, 0 < (vz-yz) < 0.7),$$

the relation for dE,N galaxies is very similar:

$$[Fe/H] = -2.83 + 3.78(vz-yz)(0.4 < (vz-yz) < 0.7).$$

In general, E galaxies show real deviation of less than ±0.5 dex in the metallicity according the individual history of stellar populations especially in clusters environment.

Our future plans are to observe dwarf elliptical galaxies in galaxy clusters with higher redshifts, to follow the behavior of these galaxies closer to the time of their formation. Success will require that we reach galaxies as faint as M= −16 and, thus, the need for extragalactic research on VLT telescopes.

*K. Rakos, J. Schombert, A. Odell, M. Maitzen*

### 1.3 Project Goals

For the first time the dwarf elliptical galaxies (selected by photometry) can be observed at redshifts greater than 0.4 which corresponds to 4 Gyrs younger than today. With future data, we will be able to test the correlation between density of galaxy clusters and number of dE galaxies (residuals of destroyed galaxies), the correlation between age of dE,N galaxies and their metallicity and the correlation between the density of galaxy clusters and the metallicity of dE,N galaxies in the clusters.

The authors wish to thank NOAO and Steward Observatory for granting telescope time for this project. One of us (K. Rakos) gratefully acknowledges the financial support from the Austrian Fonds zur Foerderung der wissenschaftlichen Forschung.